\documentclass[preprint2]{aastex}

\shorttitle{Lithium in halo dwarfs}
\shortauthors{L. Piau}

\begin{document}


\title{Lithium isotopes in population II dwarfs}


\author{L. Piau\altaffilmark{1}}
\affil{University of Chicago, LASR, 933, East 56th street, Chicago, Il, 60637, USA}



\begin{abstract}
We investigate the history of $^6$Li and $^7$Li in population II dwarfs
during the pre main sequence and main sequence. The evolution is followed using the
CESAM code and taking into account the most recent physics.
The effective temperature 
ranges from $\approx$ 4700 K to $\approx$ 6400 K and therefore concerns objects on the 
so-called Spite plateau and cooler. 
We find the $^7$Li pre main sequence depletion is unable to account for the observations
in the halo whatever the effective temperature. This supports microscopic diffusion 
and an additional non standard
mixing process both acting during the main sequence. On the contrary the models $^6$Li pre main 
sequence depletion appears too strong and is marginally compatible with recent detections.

During the main sequence we introduce the effects of tachocline diffusion. 
This process is a rotationnally induced mixing acting at the top of the radiative core.
We show that the differences in the early 
rotation history cannot result in scattered lithium abundances on the Spite plateau.
Moreover the tachocline mixing process predicts $^7$Li abundances in good agreement
with the observations.
We briefly address the question of turn off $^7$Li poor stars.
A modest accretion of lithium free matter 
would be enough to explain their low abundance in this element.

We expect the scatter in $^7$Li abundances is correlated to variations in [Fe/O] 
ratio for dwarfs cooler than  5500 K.
Finally the tachocline mixing is robust with respect to the recent $^6$Li observations
around the turn-off. We similarly suggest the [Fe/O] should be higher in objects with
$^6$Li and effective temperature below 6000 K.
\end{abstract}





\section{Introduction}\label{sec1}

Lithium, beryllium, and boron isotopes get destroyed by proton capture
in stellar interiors between 2 $\rm 10^6$ K and 5 $\rm 10^6$ K.
Therefore their surface abundances allow direct measurement of the depth of the 
outer convection zones. In addition when the temperature
near the base of a convection zone (hereafter BCZ) is too low 
they permit to set constraints on any process that 
would mix the matter beneath. The light elements
have been extensively observed in main 
sequence low mass stars. With asteroseismology they are currently 
the only means to constrain low mass dwarfs' interiors.

Population II main sequence stars being the oldest non-evolved objects, they
should reflect the initial chemical conditions
of our Galaxy. It is therefore particularly interesting to
set constraints on their possible surface abundance changes since they were formed.
In particular $\rm ^7Li$ has been extensively studied in these objects 
both from theoretical and observational viewpoints. 
$\rm ^7Li$ is one of the rare nucleids that
was synthetised through Big Bang Nucleosynthesis (hereafter BBN)
in an observable amount. 
Estimating its initial abundance from 
observations of the oldest stars provides essential indications on the
conditions in the primordial Universe.
More than 20 years ago Spite \& Spite (1982) found that most 
halo field stars with $\rm [Fe/H]<-1.5 dex$ and $\rm  5500 K < T_{eff} < 6300 K$ display similar
$^7\rm Li$ abundances. Since then this result was confirmed by many
other observations (Hobbs \& Duncan 1987; 
Spite \& Spite 1993; Thorburn 1994; Ryan et al. 1996). 
The $\rm ^7Li$ abundance on the so-called Spite plateau
is now estimated to lie around 2.0-2.2 dex with a very small intrinsic 
scatter below 0.1 dex. The key issue for stellar physicists 
is the determination of how this abundance has changed since the halo
stars were formed.

Recently, new spectrographs such as UVES on the VLT have 
considerably increased the number of 
measurements in other light elements isotopes such as
$^6\rm Li$ (Asplund et al 2005a, 2005b) and $^9\rm Be$. This makes the 
determination of their evolution from the modelling point of view relevant.
In contrast to $^7\rm Li$, these nucleids
are not produced in a significant manner through BBN but are the
result of fusion or spallation processes in the interstellar medium 
as suggested by the pioneering work of Reeves et al. (1970).
Beside this they have different depletion temperatures than $\rm ^7Li$ so that
they provide complementary information on stellar structure and evolution.
$^6$Li is especially intriguing as present measurements suggest it does
not evolve over a broad range of metallicity
(Asplund et al. 2005a) whereas it is expected to increase substantialy due to 
production by low energy cosmic rays (Vangioni-Flam et al. 1999).

In this work we present the predictions of models of population II low mass
stars on the $\rm ^6Li$ and $\rm ^7Li$ surface abundances. The models
are evolved up to 13 Gyr, a plausible age for halo stars having 
[Fe/H]=-2 dex (VandenBerg 2000). The history 
of the isotopes is investigated both on the pre main sequence (hereafter pre MS) 
and the main sequence (hereafter MS) using the most recent physics in terms 
of nuclear reaction rates, equation of state and opacities. We have, moreover,
studied the effects of the rotationally induced tachocline mixing.
This process has given encouraging results for the evolution of light elements 
on the surface of population I solar analogs but has never been studied 
in population II objects. The paper is organized as follows :
in \S \ref{sec2} we describe in detail the general inputs to our code
and the assumed initial composition. The rotation history has been
followed in detail. In \S \ref{sec3} we 
present our assumptions on the rotation history and consequently on the 
tachocline mixing process. Sections \ref{sec4} and \ref{sec5} respectively investigate the
effects of pre MS and MS history on $\rm ^6Li$ and $\rm ^7Li$.
Finally we discuss our results in \S \ref{sec6}.

\section{The evolutionary code and general inputs}\label{sec2}

We use the CESAM code to perform the computations (Morel 1997).
This hydrostatic onedimensional code has been 
extensively employed to model various stellar types and 
stages of evolution over the last decade. It has also been used in solar structure computations 
and helio-/asteroseismology\footnote{The list of the scientific publications using CESAM is available 
at http://www.obs-nice.fr/cesam/}.
Various adaptations have been made to the code in order to use the most adequate
inputs for the present issue. 

{\it The equation of state}: in the case of population
II objects the metals' impact on the equation of state (EOS) is negligible. 
Consequently we use the OPAL2001 EOS tables for pure hydrogen-helium mixtures 
\footnote{Available at http://www-phys.llnl.gov/Research/OPAL/} 
(Rogers \& Nayfonov 2002).

{\it The opacities}: consistent with the EOS, we use the OPAL opacities.
The opacity table considered is the $\alpha$-element enhanced
table of F. Allard. For this table $\rm [\alpha/Fe]=0.3 dex$
where $\rm \alpha$ stands for the $\rm \alpha$-elements (see hereafter). 
Low temperature opacity tables are based on computations similar to
Alexander and Ferguson (1994). Below log T =3.75
the OPAL opacity tables are replaced by the tables provided by J. Ferguson (private communication).
These tables were generated for the same intermetallic
ratios as those of the OPAL opacity tables.

{\it The nuclear reactions}: we use the NACRE compilation of nuclear reaction rates  
(Angulo et al. 1999). Our network includes the proton-proton chains 
and CNO cycle. Furthermore the $\rm ^6Li(p,\alpha)^3He$  and $\rm ^9Be(p,^2H)2^4He$ 
reactions are taken into account.

{\it The atmosphere modelling}: the outer boundary conditions to the stellar
structure equations are provided using the Nextgen atmosphere models.
The atmospheric temperature-optical depth relations have kindly been computed by P. Hauschildt 
for the required composition and on a grid that spans the effective temperature-surface
gravity domain we encounter. 

{\it The convection}: the convection zones are assumed to be fully
homogeneous, and modeled using the mixing-length theory
(hereafter MLT) in a formalism almost identical to that of B\"ohm-Vitense (1958).
This formalism is precisely described in the appendix of Piau et al. (2005).
We have considered $\rm \alpha_{MLT}=1.766$ which is our Sun-calibrated value.

{\it The diffusion processes}: The microscopic
diffusion is taken into account following the Michaud \& Proffitt
(1993) prescriptions. The details of the 
tachocline diffusion process are given in \S \ref{sec3}.

In addition to these different points the initial composition 
is an essential question. For instance, the impact of metallicity on the  
extension of convection during the pre MS is a priori unknown. This impact is strong
in the case of population I object (Piau \& Turck-Chi\`eze 2002, hereafter PTC02). 
Moreover in MS models changing the intermetallic ratios affects 
the temperature at the base of the convection zone for a given effective 
temperature. Both phenomona could significantly change the light elements' histories.
We assume [Fe/H]=-2.0 dex and [$\alpha$/Fe]=+0.3 dex. 
$\alpha$ stands for the oxygen and the usual $\alpha$-elements 
(Ne,Mg,Si,S,Ar,Ca and Ti). We consider a similar ratio for nickel and 
chromium as for iron while manganese and aluminum are supposed underabundant 
by 0.15 dex and 0.30 dex with respect to iron (Gratton 1989; Magain 1989).
Recent observations of the [C/O] vs [O/H]
function in halo stars (Akerman et al. 2004) suggest that the carbon is underabundant with respect to 
oxygen there. The deficiency being around 0.4 dex for the oxygen fraction accounted in our models
it would bring the carbon to iron ratio close to solar. We have thus considered [C/Fe]=0 which
is also in agreement with other studies regarding carbon in population II (McWilliam 1995).
\footnote{The situation seems different in extremly metal
poor stars ($\rm -4.1 < [Fe/H] < -2.7 dex$) where [C/Fe] or [(C+N)/Fe] is close to 0.2 dex 
(Cayrel et al. 2004, Spite et al. 2005).
However, only the objects with [Fe/H] around -2.0 dex are within the scope of this work.}
Nitrogen is assumed to follow the same trend as carbon so
that [N/Fe]=0. The total helium mass fraction along with the 
ratios of $\rm ^2H$, $\rm ^3He$, and $\rm ^7Li$ to $^1$H
are set to their primordial values following the recent BBN 
calculations of Coc et al. (2004). We consider that
$^6$Li=1.13 dex and $^9$Be=-0.15 dex initially which is adapted 
from the observations of turn-off population II 
stars by Cayrel et al. (1999) and Pasquini et al. (2004) respectively.
In both last cases we increase the observed abundances by 0.2 dex in order to
compensate for the microscopic diffusion at the halo age.
$^9$Be surface abundance is followed,
however because of the higher temperature for $^9$Be proton
capture it is never affected by nuclear reactions on our range
of $\rm T_{eff}$. We therefore shall not discuss $^9$Be any further
in this study.
The tables \ref{tab1} and \ref{tab2} provide the detailed composition.
Hydrogen and helium mass fractions
are X=0.7518 and Y=0.2479 respectively. Unless explicitly
mentioned we always adopt this composition.

\begin{table*}[ht]
  \begin{center}
    \caption{Metal fractions relative to the Sun.}\vspace{1em}
    \renewcommand{\arraystretch}{1.2}
    \begin{tabular}[h]{lc}
      	Metals                        &  [X/H]  \\
      \hline
	O,Ne,Na,Mg,Si,P,S,Cl,Ar,Ca,Ti & -1.7    \\
      \hline
	Fe,Ni,Cr,C,N,K                & -2.0    \\
      \hline
	Mn                            & -2.15   \\
      \hline
	Al                            & -2.30   \\
      \hline
      \end{tabular}
   \label{tab1}
  \end{center}
\end{table*}

\begin{table*}[ht]
  \begin{center}
    \caption{Light element number abundances relative to hydrogen adapted from  Coc et. al (2004)
 except for $^6$Li and $^9$Be (see text). Following Coc et. al (2004) we consider
            the helium mass fraction to be Y=0.2479.}\vspace{1em}
    \renewcommand{\arraystretch}{1.2}
    \begin{tabular}[h]{lccccc}
     ratios & $\rm ^2H / ^1H$ & $\rm ^3He / ^1H$  &  $\rm ^6Li / ^1H$  &  $\rm ^7Li / ^1H$  & $\rm ^9Be / ^1H$   \\
      \hline
     values & $2.60\,10^{-5}$ & $1.04\,10^{-5}$        &  $1.34\,10^{-11}$  & $4.15\,10^{-10}$   & $7.07\,10^{-13}$   \\
      \hline
     values (log) & -4.58     & -4.98                  &    1.13              & 2.62                & -0.15   \\
      \hline
      \end{tabular}
   \label{tab2}
  \end{center}
\end{table*}

\section{The rotation}\label{sec3}

Population II stars cannot be observed
during their evolution.
Thus it is impossible to observe their surface rotation history.
Currently all the main sequence population II stars exhibit very low equatorial
velocities. The line broadening due to these rotation
rates is comparable to the spectrographs resolution today and for most objects
only upper limits of the projected equatorial velocities are accessible. 
Studying 9 halo stars Smith et al. (1998) found $\rm v sini < 3\,km.s^{-1}$ for all
of them. Even this upper limit was lowered to 2 km/s by more 
recent observations (see Fig. 1 of Ryan et al. 2002).
Interestingly, the latter authors also report that 3 stars out of their sample
exhibit slightly higher $\rm v sin i$ in the 5.5 to 8.3 $\rm km.s^{-1}$ range. 
These peculiar stars are spectroscopic binaries.
They exhibit $\rm ^7Li$ surface abundances at least 0.5 dex below the other objects
and should not be considered when studying the isolated stars general
lithium history we address here. However we briefly touch on the 
question of such Li-poor objects in \S \ref{sec5}.

We compute the rotation evolution by applying different
assumptions to the angular momentum losses and internal transport for a 
$\rm 0.7 M_{\odot}$ model. 
With $\rm T_{eff}=5890 K$ at 13 Gyr this star lies in the middle of the Li-plateau.
Four different models of rotation were built.
In the following subsections we present our assumptions on the rotation
history. First we discuss our choices of surface angular momentum
loss laws that in turn determine the surface rotation. 
Second we address the question of the inner 
radiation core rotation that is the relevant quantity
for the tachocline diffusion efficiency. Third we briefly 
describe the way tachocline mixing was implemented.

\subsection{The surface rotation}

We make the assumption that population II low mass
stars experience an evolution of surface rotation which is similar 
to their population I counterparts:

i) The star initially rotates as a solid body and
following Bouvier et al. (1997) (hereafter BFA97) we consider
an initial rotational period of $\rm P_o=$8 days. This period remains 
unchanged until the star decouples from its initial
accretion disk which we assume occurs at $\rm \tau_d=$3 Myr (case A) 
or 0.5 Myr (case B).

ii) Once the star is no longer locked to its disk 
the angular momentum decreases following the Kawaler (1988) prescription. 

\begin{equation}
\frac{dJ}{dt}=-K{\Omega}^3({\frac{R}{R_{\odot}}})^{1/2}({\frac{M}{M_{\odot}}})^{-1/2}\,\rm if \, \Omega
< {\Omega}_{sat}
\label{eq1}
\end{equation}

\begin{equation}
\frac{dJ}{dt}=-K{\Omega}{{\Omega}_{sat}}^2({\frac{R}{R_{\odot}}})^{1/2}({\frac{M}{M_{\odot}}})^{-1/
2}\,\rm if \, \Omega > {\Omega}_{sat}
\label{eq2}
\end{equation}

K is set to $\rm K_{\odot}=3.25 \, 10^{47} g.s.cm^2$ (cases A and B) which 
leads to the actual solar rotation for a solar model. For case C we take
$\rm K=K_{\odot}/10$. This second value
yields a $\rm 1.6\, km.s^{-1}$ equatorial velocity at 13 Gyr
which is nearly the maximum velocity allowed by current observations ($\rm v sin i < 2 km.s^{-1}$).
The surface magnetic activity saturates around $\rm 10 \Omega_{\odot}$.
The actual threshold $\rm \Omega_{sat}=14 \Omega_{\odot}$ ($\rm 3.78\, 10^{-5} rad.s^{-1}$)
is accurately tuned from the rotation observations
in middle aged open-clusters (BFA97).

iii) To compute the rotational rate inside the
radiation zone we explore two possibilities regarding its
coupling to the convective envelope.
Firstly we consider that these regions are instantaneously coupled
so that the whole star rotates as a solid always (cases A, B, C).
Secondly we consider the radiative and convective
zones to rotate as solids but not necessarily with the same speeds 
(case D, see the next subsection).

\begin{figure}[Ht]
\centering
\includegraphics[angle=90,width=8cm]{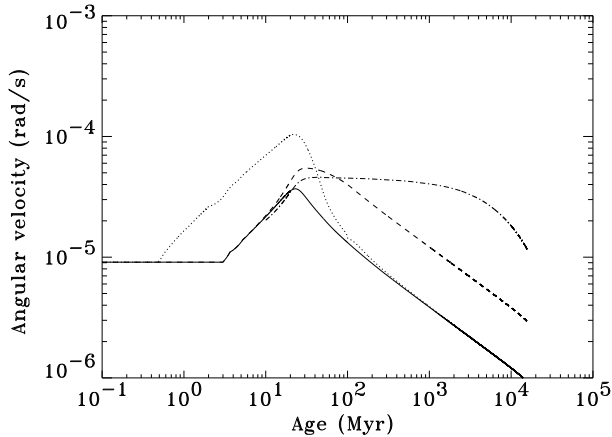}
\caption{Radiation zone angular velocity vs age for a $\rm 0.7 M_{\odot}$ 
model. The solid body
rotation models are shown by the solid line : $\rm K=K_{\odot}$, and  $\rm \tau_d=3$ Myr (case A);
the dotted line :  $\rm K=K_{\odot}$, and  $\rm \tau_d=0.5$ Myr (case B);
the dashed line :  $\rm K=K_{\odot}/10$, and  $\rm \tau_d=3$ Myr (case C).
Dot-dashed line represents the differential rotation model with
$\rm K=K_{\odot}$, $\rm \tau_d=3$ Myr and
$\rm \tau_c=1$ Gyr (case D). The difference between the case A
and case B models is related to early conditions and vanishes 
after $\rm \sim$ 500 Myr. }
\label{fig1} 
\end{figure}

The four empirical parameters ($\rm P_o$,$\rm \tau_d$, $\rm \Omega_{sat}$,K) in the rotation law
we adopt are directly calibrated from the observations on our Sun and population I solar analogs.
It seems reasonable to assume similar angular momentum losses
through magnetic braking for solar analogs and Li-plateau stars.
The reason is that, in terms of surface conditions
or convection zone extension, the Sun is quite comparable to
our $\rm 0.7 M_{\odot}$ population II [Fe/H]=-2 dex model.
At 13 Gyr we find for this object : $\rm T_{eff}=5890$ K, surface gravity = 1.36 $\rm g_{\odot}$
and 0.764 $\rm R_{\star}$ as radius for the radiative core vs 0.713 $\rm R_{\odot}$ for the actual Sun
(global properties of plateau stars as a function of their masses 
can be found in Proffitt \& Michaud 1991).
We consequently apply the braking law parameters from BFA97.
The shorter $\rm P_o$ and/or $\rm \tau_d$ the higher zero age main sequence 
(hereafter ZAMS) rotation velocity (figure \ref{fig1}). 
Then a higher $\rm \Omega_{sat}$ implies a slower pace for
rotational slow-down during early MS. To study the effects of
fast ZAMS rotation we built a model with $\rm \tau_d=0.5$ Myr. 
We stress however that none of the
three parameters ($\rm P_o$, $\rm \tau_d$ and $\rm \Omega_{sat}$) has a 
significant impact on rotation 
after $\sim 0.5$ Gyr where, in contrast, the K parameter plays a major role.
K determines the star's asymptotic velocity. Considering  $\rm \tau_d=3$ Myr, 
we explore its impact when it is varied from $\rm K_{\odot}$ to $\rm K_{\odot}/10$ 
for which it causes surface
rotation velocities to vary from $\rm \sim 0.6$ to $\rm \sim 1.7 km.s^{-1}$ 
at 13 Gyr (figure \ref{fig1}). 

\subsection{The internal rotation}

The rotational rate in the upper radiation zone 
is the relevant quantity for the tachocline diffusion computation. 
This rate is also the surface rotation rate when the rotation is solid.
In the case of population I solar like stars there is an accumulation of observational and
theoretical clues suggesting that this should be correct during most of
the main sequence.
Helioseismic observations show that our Sun
rotates nearly as a solid at least down to $\rm 0.2 R_{\odot}$
(Thompson et al. 2003). As shown in Piau et al. (2003), 
the surface rotation history in the
young clusters does not seem compatible with differential
rotation with the radius.
In this last work we came to the conclusion that solid
body rotation was certainly achieved as early as the Hyades age.
Furthermore one would
expect more significant surface magnetic activity in young
slow rotators if they had fast spinning cores (BFA97). Finally 
some recent models of angular momentum transport by internal waves 
suggest an efficient rigidification of the stellar interior both 
for the Sun (Talon, Kumar \& Zahn 2002) and for 
the lithium plateau stars (Talon \& Charbonnel 2004).

In the case of lithium plateau stars however we cannot completely
exclude differential rotation with depth because part of the low metallicity 
horizontal branch (hereafter HB) stars show significant
surface rotation while on the other hand the turn off halo stars 
are extremely slow rotators. We therefore built a rotation
model assuming $\rm K= K_{\odot}$, $\rm \tau_d=3$ Myr and
$\tau_c=1$ Gyr. For this model, during the time interval dt, 
the radiative core and the convective envelope
exchange an angular momentum fraction $\rm dJ=\frac{\Delta J}{\tau_c}dt$.
$$\rm \Delta J =\frac{(I_{env}J_{core} - I_{core}J_{env})}{I_{core} + I_{env}}$$
is the angular momentum exchange 
that would achieve synchronisation of the core and the envelope, 
$\rm \tau_c$ is therefore a coupling timescale for rotation 
between these regions. As shown on figure
\ref{fig1} 
this option yields to a significant increase of the rotation
rate in the radiation zone. At 13 Gyr 
the core of the $\rm 0.7 \, M_{\odot}$ star rotates at $\rm 1.35\,10^{-5}\,rad.s^{-1}$ while the
surface rotation velocity linear is $\rm 1.5 \, km.s^{-1}$.

For this model the radiative core rotates $\rm \sim 5$ times
faster than the envelope. Indeed this high
ratio could be expected from constraints on fast rotating HB stars.
The coolest HB stars ($\rm T_{eff} < 12000 K$) are spinning as fast as 
$\rm 40 km.s^{-1}$ (Behr 2003). Sills and Pinsonneault (2000) made a systematic
study of the angular momentum history during the red giant branch evolution.
Their models save the greatest amount of angular momentum when specific angular
momentum is constant in the convective envelope and
fully retained in the radiative core. For this situation 
the authors demonstrated that the fast rotating HB stars could be accounted for 
if turn-off progenitors rotate at $\rm 4 km.s^{-1}$.
The subsequent observations of Ryan et al. (2002)
suggest $\rm v sin i < 2 km/.^{-1}$ however. Thus if these turn-off stars rotate
as solid bodies their low angular momentum seems difficult to reconcile
with the high rotation velocities of the HB stars even for
very favorable conditions of angular momentum distribution.
We can estimate the rotational velocity of the core
for a $\rm 0.8 M_{\odot}$ turnoff star under the following hypotheses:
i) The total angular momentum of the star is similar to its HB value
: $\rm J_{HB}=J_{core}+J_{env}$.
$\rm J_{HB}$, $\rm J_{core}$ and $\rm J_{env}$ respectively stand for the angular momenta of all the
HB star, its turn off progenitor core and envelope.
ii) Both the whole HB star and 
the radiative core of its turn off progenitor rotate as solids.
iii) The surface velocities of the HB star and its turn off progenitor
are respectively $\rm 40 km.s^{-1}$ and $\rm 2 km.s^{-1}$. 

Then
\begin{equation}
\rm \Omega_{core}=\frac{I_{HB}\Omega_{HB}-I_{env}\Omega_{env}}{I_{core}}
\label{eq3}
\end{equation}
 
At 11.8 Gyr when central hydrogen exhaustion occurs ($\rm X_c=10^{-3}$)
$\rm I_{core}=4.08\,10^{53}\,g.cm^2$, $\rm I_{env}=3.71\,10^{51}\,g.cm^2$, and
$\rm \Omega_{env}=2.45\,10^{-6}rad.s^{-1}$. The rotation rate of the core 
is thus $\rm \Omega_{core}=3.38 10^{-5} rad.s^{-1}$.
Now if we compute the rotational evolution of this $\rm 0.8 M_{\odot}$ model
with the case D parameters we find  $\rm \Omega_{core}=4.7 10^{-5} rad.s^{-1}$ at 
the same age. Both evaluations are quite comparable.

Before moving to the rotation effects in the next sections 
let us sum up our four different assumptions on rotation 
and what motivates them :

Case A): Solid body rotation, $\rm K=K_{\odot}$, and  $\rm \tau_d=3$ Myr.
This rotation model is calibrated thanks to the Sun and population I observations.
Unless explicitly mentioned we will always use this rotational law.

Case B): Solid body rotation, $\rm K=K_{\odot}$, and  $\rm \tau_d=0.5$ Myr.

Case C): Solid body rotation, $\rm K=K_{\odot}/10$, and  $\rm \tau_d=3$ Myr.
Cases B) and C) evaluate the impact of
increased rotation rates with respect to  case A) 
either because of lower angular momentum losses
or higher ZAMS rotation speed. 

Case D): Differential rotation model $\rm K= K_{\odot}$, and  $\rm \tau_d=3$ Myr.
The coupling time scale between the convective envelope and
the radiative core is $\rm \tau_c=1$ Gyr. This model
aims at evaluating the effects of a possible fast rotating core.

\subsection{The tachocline mixing}

The transition region between the solar convective and radiative zones shows
sharp variations of the rotation velocity with depth and has therefore received the name
of tachocline. There rotation and differential rotation with
latitude induce a slow mixing (Spiegel \& Zahn 1992).
The tachocline mixing has been introduced in modelling studies
about the Sun and population I solar-like stars (Brun et al. 1999, 2002; Piau et al. 2003). 
It produces better agreement between theoretical and observed sound speed
and explains the $\rm ^7Li$ and $\rm ^9Be$ main sequence histories.
However, the tachocline mixing effect has never been systematically studied in population II stars.

We presented in detail our assumptions on the global rotation history
hereabove. The differential rotation $\tilde{\Omega}$ with latitude in the convection zone 
is then deduced 
assuming it is the solar differential rotation for the solar 
rotation rate and otherwise scales as $\Omega^{0.7}$ (Donahue, 
Saar \& Baliunas 1996). It is noteworthy that
the tachocline mixing has features making it robust with respect 
to the present constraints on stellar rotation: it is not related to differential
rotation with depth which presently seems excluded in MS stars by observations as well
as by new models of internal angular momentum transport. Besides this and
contrary to models where angular momentum is extracted by a diffusive process, 
it infers no $^9$Be depletion in solar analogs. Our tachocline
mixing prescription has two free parameters that must be empirically calibrated: the 
width of the tachocline and the
buoyancy frequency in the tachocline region. The width is calibrated from solar seismic measurements
to 2.5\% of the total radius.
The value of the buoyancy frequency is estimated to be 10 $\rm \mu$Hz. A change 
to 2 $\rm \mu$Hz is briefly investigated however (figure \ref{fig4}).
A discussion of the choice of these values can be found
in Piau et al. (2003).

We do not take into account the tachocline 
mixing during the pre main sequence (which is considered to end
at 200 Myr) unless explicitly mentioned. The reason
is that the time for effective diffusion induced by the tachocline
process is the order of : $$\rm t_{diff}=450(\frac{\tilde{\Omega}}{\Omega})^{-1} Myr$$
(Zahn 2004). Even in the case of extreme differential
rotation ($\tilde{\Omega}=\Omega$) this duration thus already exceeds 200 Myr.
A detailed description of the tachocline process equations
can be found in Zahn (2004).

\section{Lithium isotopes during the pre main sequence}\label{sec4}

\begin{figure}[Ht]
\centering
\includegraphics[angle=90,width=8cm]{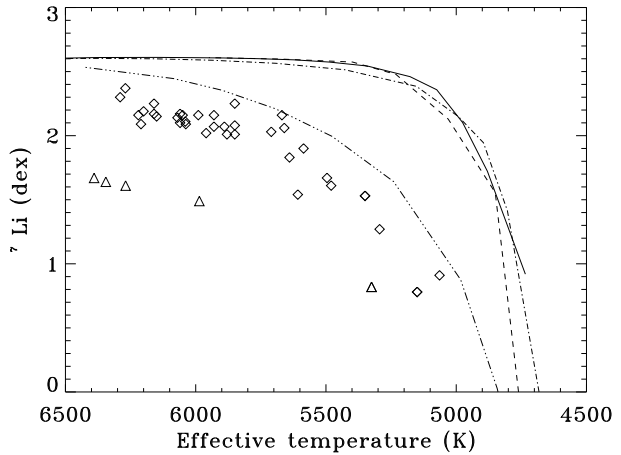}
\caption{$^7$Li pre MS depletion at 200 Myr in [Fe/H]=-2 dex (solid line) , [Fe/H]=-1 dex (dashed line), [Fe/H]=-3 dex
(dash-dotted line) models. These models all exhibit similar pre MS $^7$Li depletion patterns. The [Fe/H]=-0.7 dex models
(dash three dotted line) exhibit higher depletion because of the increasing impact of metals on opacity.
Observations (diamonds) and upper limits (triangles) are from Ryan et al. (1996), Ryan \& Deliyannis (1998), 
Ryan et al. (1999) and  Ryan et al. (2001). For the sake of clarity error bars are not shown.}
\label{fig2} 
\end{figure}

During the pre MS, low mass stars presumably evolve
from a fully convective state to a radiative core and convective envelope structure.
The temperature at the BCZ reaches a maximum which may induce light element depletion
visible at the surface (see PTC02 and references therein). The lower 
the mass (i.e. the lower the ZAMS $\rm T_{eff}$), the deeper the convection zone at 
any stage of the evolution and the stronger the light element depletion.
We evaluate the pre MS depletion of $^6$Li and $^7$Li based
on models including no tachocline diffusion. The impact of the pre MS 
evolution is considered at 200 Myr. At this age
the proton-proton chains provide more than 99\% of 
the energy while the microscopic
diffusion is negligible for all our models.
We consider therefore 200 Myr as the end of pre MS. The computations were made for
the helium mass fraction Y=0.2479 and the metallicities [Fe/H]=-3 dex, [Fe/H]=-2 dex (standard case), 
[Fe/H]=-1 dex and [Fe/H]=-0.7 dex. 
As shown in table \ref{tab3} and figure \ref{fig2} no significant $\rm ^7Li$ depletion is 
induced during pre MS even for objects of ZAMS effective temperature at
the cool end of the Spite plateau ($\rm \sim 0.65 M_{\odot}$). Below this limit 
we estimate a depletion phenomenon rapidly increasing as the effective temperature
decreases. These predictions are however unable to fully explains 
the $\rm T_{eff}- ^7Li$ pattern presently observed below 5500 K. Firstly in the $\rm T_{eff}-^7Li$ plan,
we compute a 0.18 dex $\rm ^7Li$ decrease per 100 K between 5175 K 
($\rm 0.62 M_{\odot}$ model) and 4975 K ($\rm 0.58 M_{\odot}$ model).
This slope is smaller than the estimate of 0.27 dex based on 
observations between 5000 K and 5500 K suggested by 
Ryan \& Deliyannis (1998). Second the predicted depletion rate around 5000 K is clearly lower
than what is observed.
Finally even the current intensity computed for pre MS $\rm ^7Li$ depletion could be
questioned: in the case of the population I G-type stars it 
clearly exceeds what is observed (Ventura et al. 1998; PTC02).  
We conclude that the $\rm T_{eff}-^7Li$ relation currently observed in the halo
stars and for temperatures cooler than the cool end of the Spite plateau 
was (at least partially) built after the pre MS. 

\begin{table*}[ht]
  \begin{center}
    \caption{ZAMS (age = 200 Myr)lithium to initial lithium ratios as function of temperature
for microscopic diffusion models : [Fe/H]=-2 dex and Y=0.2479.}\vspace{1em}
    \renewcommand{\arraystretch}{1.2}
    \begin{tabular}[h]{lcccccc}
      Mass ($\rm M_{\odot}$)        &  0.80          & 0.75         &   0.70         & 0.65           &  0.62    &  0.58       \\
      \hline  
     $\rm ZAMS \, T_{eff} (K)$      &  6144          & 5893         &   5621         & 5341           &  5177    &  4975       \\
      \hline
     $\rm ^7Li_{ZAMS}/^7Li_{0}$     &  0.97          & 0.96         &   0.93         & 0.83           &  0.69    &  0.30       \\
      \hline
     $\rm ^6Li_{ZAMS}/^6Li_{0}$     &  0.43          & 0.15         & $1.1\,10^{-2}$ & $3.6 \,10^{-6}$& $<10^{-7}$& $<10^{-7}$ \\
      \hline
      \end{tabular}
   \label{tab3}
  \end{center}
\end{table*}

It is interesting to note that the computed pre MS $^7$Li depletion does 
not depend on the metallicity from [Fe/H]=-3 to [Fe/H]=-1 dex. 
As shown on figure \ref{fig2} the predicted $\rm ^7Li-T_{eff}$ relations are almost identical
on this range of metal fraction. This point is a marked 
difference from the pre MS population I depletion predictions.
Contrary to population I objects the metals represent a very small fraction 
of the plasma components in the halo dwarfs. Despite the numerous bound-bound and bound-free interactions
with the radiation field the contribution of heavy chemical species 
to the opacity budget near the BCZ becomes small below [Fe/H]=-1dex. We computed the metal 
contribution to the opacity in the [Fe/H]=-0.7, -1 and -2 dex cases. 
For this purpose we used the monochromatic calculations of Iglesias \& Rogers (1996).
The contribution of a component is computed by subtracting
the opacity evaluated for the mixture without the component to the opacity of the
global mixture. This requires that the component whose opacity contribution is evaluated 
does not provide a significantly fraction of the free electrons\footnote{In this
respect we remark that the evaluation of element opacities Table 3 of Piau \& Turck-Chi\`eze 
(2002) is incorrect in the very cases of hydrogen and helium.}.
Table \ref{tab4} compares these computations showing the increasing role of metals
in the opacity budget with increasing [Fe/H].

\begin{table*}[ht]
  \begin{center}
    \caption{Opacity and impact of the main metals contributing to the opacity (O, Ne, Mg, Si, S and Fe) at the BCZ. 
The opacities are computed at the peaks of BCZ temperature (i.e maximum lithium depletion rate) in three objects 
that end up with similar effective temperatures on the ZAMS. The ZAMS effective temperatures are 4975 K, 4991 K and 
4982 K for the [Fe/H]=-2 dex, -1 dex and -0.7 dex respectively.
This temperature corresponds to a moderate $^7$Li depletion during the pre-MS (see fig \ref{fig2}).}\vspace{1em}
    \renewcommand{\arraystretch}{1.2}
    \begin{tabular}[h]{lcccccc}
               &    [Fe/H]=-2    &   [Fe/H]=-1    &   [Fe/H]=-0.7   \\
               &    $\rm M_{\star}=0.58 M_{\odot}$   &  $\rm M_{\star}=0.62 M_{\odot}$    &   $\rm M_{\star}=0.70 M_{\odot}$  \\
      \hline 
      \hline  
 Pre MS maximal BCZ temperature  &          $\rm 3.44 \, 10^6 K$     &            $\rm 3.45 \, 10^6 K$                         & $\rm 3.79 \, 10^6 K$     \\
 \& corresponding density        &          $\rm 3.98 g.cm^{-3}$  &            $\rm 3.95 g.cm^{-3}$                      & $\rm 2.66 g.cm^{-3}$  \\
      \hline  
     Opacity  ($\rm \kappa$)  &          5.37 $\rm cm^{2}.g^{-1}$ &            8.95 $\rm cm^{2}.g^{-1}$                  &           7.62  $\rm cm^{2}.g^{-1}$       \\
      \hline
   Contribution of & & & \\
   O Ne Mg Si S Fe to $\rm \kappa$  &     7.9 \%                       &      43 \%                                 &             61 \%                               \\
      \hline
   Contribution of & & & \\
   C N Na Al P Cl Ar & & & \\
   K Ca Ti Cr Mn Ni to $\rm \kappa$  &    0.4 \%                        &      2.6 \%                                      &     3.9 \%                             \\
      \hline
      \end{tabular}
   \label{tab4}
  \end{center}
\end{table*}

\begin{figure}[Ht]
\centering
\includegraphics[angle=90,width=8cm]{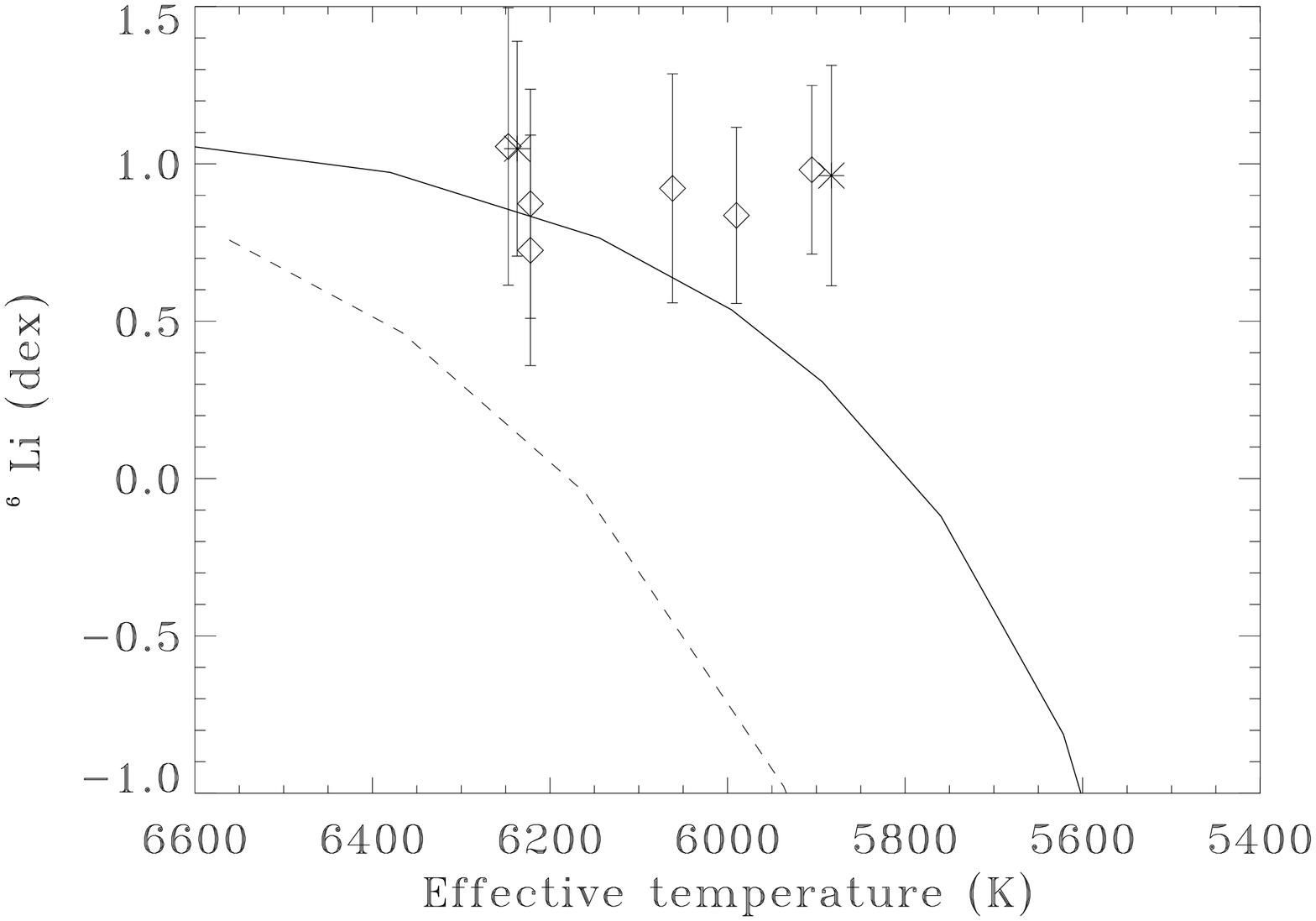}
\caption{$^6$Li pre MS depletion in [Fe/H]=-2 dex (solid line) and [Fe/H]=-1 dex (dashed line).
Each curve has been shifted upwards by 0.2 dex in order to account for
microscopic diffusion effects.
Diamonds: data in the [Fe/H]= -1 to -2 dex range from Asplund et al. (2005a) 
and M. Asplund private communication;
Stars: data from Nissen et al. (1999) $\rm [Fe/H]\approx$ -0.6 dex.}
\label{fig3} 
\end{figure}

Being more fragile, $^6$Li is more strongly affected by the pre MS peak temperatures near the BCZ.
We compare here our computations to data collected in 
Nissen et al. (1999) and mostly 
Asplund et al. (2005a, 2005b). For this sample we restrict ourselves to
the nine stars showing a non-zero $\rm ^6Li/^7Li$ 
isotopic ratio to a 2 $\rm \sigma$ confidence level. 
In order to account for the microscopic diffusion processes during MS let us recall that
our initial $^6$Li has been increased by 0.2 dex with respect to the 0.93 dex of the 
Cayrel et al. (1999) measurement in HD84937.
Our modelling of pre MS $^6$Li history 
suggests features whose consequences are clearly not observed in the present halo stars. 
First the depletion increases for decreasing effective temperature
even in the [6500,6000] K range (figure \ref{fig3}). This trend is not present in the observations,
which also systematically suggest a higher abundance than predicted.
For instance the metallicity of HD106038 ($\rm T_{eff}=5905 K$) is
[Fe/H]=-1.35 dex but this star is almost 2 dex above the $\rm T_{eff}-^6Li$ track
for [Fe/H]=-1 dex at its effective temperature. 
Also because the outer convection zones extend closer to the $^6$Li than the $^7$Li burning regions
the depletion appears to depend on the metallicity between [Fe/H]=-2 and -1 dex.
On the contrary, observed $^6$Li abundances appear independent of metallicity :
in figure \ref{fig3} the cases of
HD68284 (coolest object) and HD160617 (object closest
to 6000 K) are striking. Although they have metallicities of -0.59 dex
and -1.76 dex respectively they display similar $^6$Li fractions.
The difference we expect for a 1 dex variation of [Fe/H]
is over 1.5 dex if $\rm T_{eff} < 6000 K$. As a matter of fact
the data suggest a $^6$Li plateau with respect to $\rm [Fe/H]$ 
(Asplund 2005a; 2005b). We note that there also seems to be 
a plateau in the $\rm T_{eff}- ^6Li$ relation.
As the depletion rates are too large when compared to this plateau,
the situation of $^6$Li in population II stars is very similar
to the situation of $^7$Li in population I stars regarding pre MS: 
the actual computations predict an overdepletion (PTC02). 
Interestingly this discrepancy lingers despite very different compositions 
and the improvements in the physical inputs to the code.

\section{Lithium isotopes during the main sequence}\label{sec5}

Pre main sequence computations suggest that the 
$^7$Li abundance pattern on the Spite plateau or below $\rm T_{eff}=5500$K 
was not achieved through early evolution but during the MS. 
Furthermore $^6$Li
shows no systematic pre MS depletion at least down to $\rm \sim T_{eff}=5900$K.
In this section we study the effect of the MS tachocline mixing (Spiegel \& Zahn 1992) on
both lithium isotopes. Pre MS depletion has been systematically cancelled.
Thus we assume that all the stars exhibit on their surface 
their formation lithium on the ZAMS.
This assumption originates from the 
present computations indicating that $^6$Li 
pre MS depletion is overestimated. In order to study correctly  
the lithium isotopes history during MS one should not include pre MS 
effects that are not fully understood at the moment.  

\subsection{The $\rm ^7Li$ plateau}

When coupled to BBN computations, two independent observational facts
allow constraints on the primordial $\rm ^7Li/^1H$ ratio :
the deuterium abundance measurements 
in quasar absorption systems (Burles 2002 and references therein)
and the cosmic microwave background anisotropies probed by the 
{\it Wilkinson Microwave Anisotropy Probe} (Cyburt et al. 2003; Coc et al. 2004).
These constraints both point towards a similar $\rm ^7Li$ fraction $\sim$ 3 times 
higher than what is observed in halo dwarfs of effective temperature 
larger than 5500 K (see figure \ref{fig2}). 
The reason for this discrepancy does not currently seem to originate in
uncertainties on the BBN cross sections (Cyburt et al. 2004). The lithium depletion during 
halo dwarfs evolution therefore remains a plausible explanation.
Besides this, if one considers population I stars' histories,
a change in the surface abundances is expected for halo dwarfs.
Helioseismic measurements strongly support microscopic diffusion
in the Sun (Bahcall, Pinsonneault \& Wasserburg 1995) i.e. on a shorter timescale and for a deeper
convection zone than experienced in Li-plateau objects. Moreover, a
non standard mechanism is necessary to explain the 
time dependent $\rm ^7Li$ depletion observed in open-clusters G type stars
as well as the actual solar photospheric lithium abundance.
There is no physical reason why such a process should not be at work in population II stars.
The interplay of this process with microscopic diffusion 
sets the global lithium history.

Several mechanisms have been proposed to explain the 
$\rm ^7Li$ depletion in Galactic disk stars and were also
applied to halo stars : mass loss (Vauclair \& Charbonnel 1995), 
mixing by internal waves (Montalb\'an \& Schatzman 2000) and
rotation effects. These latter mechanisms
have been particularly explored notably as during population I MS lifetime
both depletion and rotation rates are observed to decrease. Recently Th\'eado \& Vauclair 
(2001) considered the effects of meridional circulation coupled
to molecular weight gradient effects. Their models were able to 
produce a nearly constant $\rm ^7Li$ depletion for the stars on the 
plateau without ad hoc adjustment of their parameters.
Pinsonneault et al. (2002) used a rotationaly induced mixing model
calibrated on the Sun. They found that the scatter observed on 
the plateau and the small fraction of stars below the plateau 
were compatible with a mild depletion from 0.1 to 0.2 dex.

Whatever the process responsible for the Spite plateau lithium evolution,
this process has to comply with two challenges. First
the plateau is remarkably flat with $\rm T_{eff}$: stars with quite different
outer convection zones experience similar depletion : Ryan et al.
(1996) report a $\sim 0.04$ dex $\rm ^7Li$ variation every hundred Kelvin.
Second the scatter on the plateau is extremly small for field stars. 
Spite et al. (1996) find $\rm \sigma_{\rm ^7Li} \sim 0.06-0.08$ dex,
Bonifacio \& Molaro (1997) find $\rm \sigma_{\rm ^7Li} \sim 0.07$ dex,
while Melendez \& Ramirez (2004) find $\rm \sigma_{\rm ^7Li} \sim 0.06$ dex
and Ryan et al. (1999) find $\rm \sigma_{\rm ^7Li} \leq 0.02$ dex.
Given the observational errors the scatter is in any case 
considered consistent with zero 
by the authors and clearly lies well below 0.1 dex
\footnote{The scatter observed in globular clusters such as M92
is higher (Boesgaard et al. 1998) and probably induced
by the interactions resulting from the denser stellar environment.}. 

\begin{figure}[Ht]
\centering
\includegraphics[angle=90,width=8cm]{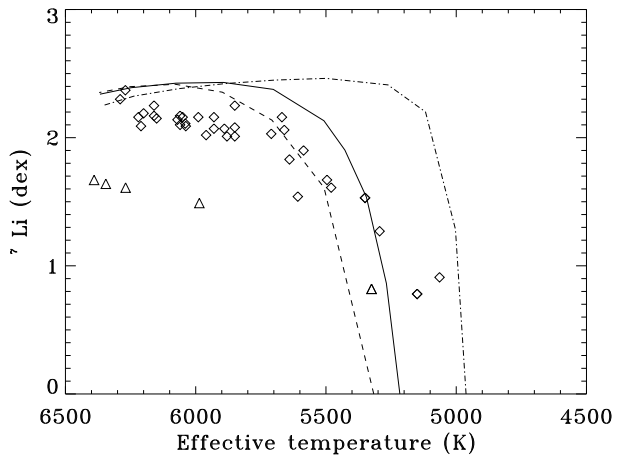}
\caption{$\rm T_{eff}-^7$Li MS relation in tachocline models of buoyancy frequency 
10 $\mu$Hz (solid line) and 2 $\mu$Hz (dashed line). The $^7$Li depletion patern 
is also provided for pure microscopic diffusion models (dot-dashed line). The 
age of the models is 13 Gyr, they all have the standard composition ([Fe/H]=-2 dex).
The data are the same as in figure \ref{fig2}.}
\label{fig4} 
\end{figure}

\begin{figure}[Ht]
\centering
\includegraphics[angle=90,width=8cm]{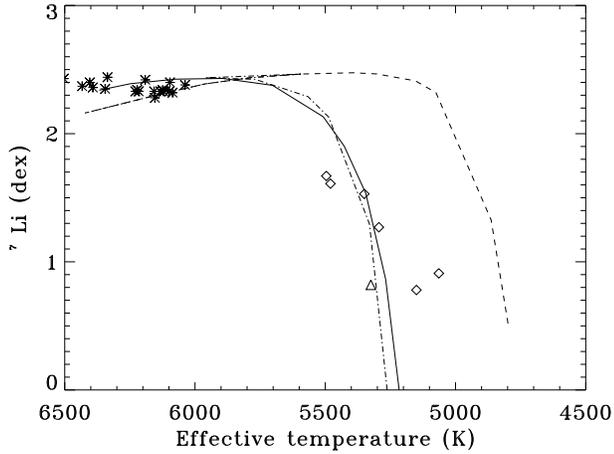}
\caption{$\rm T_{eff}-^7$Li MS relation in tachocline models of buoyancy frequency 
10 $\mu$Hz. The $^7$Li depletion pattern is provided for [Fe/H]=-2 dex (solid line),
[Fe/H]=-3 dex (dot-dashed line), [Fe/H]=-2 with the solar repartition among metals (dashed line).
The age is 13 Gyr. Note the significant impact of the metal repartition on the lithium
depletion.
The data have been taken considering the stars having $\rm -2.5<[Fe/H]<-1.5$ of the sample of 
Melendez \& Ramirez (2004) observations (star symbols) plus Ryan \& Deliyannis (1998) and Thorburn
(1994) for objects below 5500 K in $\rm T_{eff}$ (diamonds).}
\label{fig5} 
\end{figure}

We compute the MS evolution of $^7$Li for [Fe/H]=-2 dex models (composition 
described in detail in tables \ref{tab1} \& \ref{tab2}). The rotation 
law is that of case A : $\rm K=K_{\odot}$, and  $\rm \tau_d=3$ Myr
(see \S \ref{sec3}). We stress however that a different
early rotation history has no impact on the MS tachocline mixing and therefore on the
surface lithium evolution. The reason stems from the swiftly converging rotational
velocities after a few megayears combined with the few megayear
the tachocline mixing requires for its onset.
In our approach the surface abundances are not sensitive to the ZAMS angular
momentum content. In spite of a wide range of possible early rotation histories
the late MS $^7$Li contents will show no resulting scatter.
Even if we exceptionally consider the effect of the tachocline
mixing to start as early as 50 Myr,  
differences resulting from early
rotation differences (i.e. the different initial angular momentum amounts)
would not be significant. In this situation 
the difference in $^7$Li fraction we compute between
the case of a slow and a fast ZAMS rotation (case A and B respectively on figure \ref{fig1}) 
is 0.04 dex at $\rm T_{eff}=5270$ K and gets smaller above this temperature. 
Figures \ref{fig4} and \ref{fig5} present our results along with observations.
The observational data have been taken from Ryan et al. (1996),
Ryan, Norris \& Beers (1999), Ryan et al. (2001), Melendez \& Ramirez (2004)
and Charbonnel \& Primas (2005).
In order to diminish the possible $\rm ^7Li$ trends with $\rm [Fe/H]$
we selected stars with metallicity between 
-2.5 and -1.5 for the plateau objects ($\rm T_{eff} \geq 5500 K$).
Below $\rm T_{eff} = 5500 K$ we use the data from Ryan 
\& Deliyannis (1998) and Thorburn (1994). To keep a reasonable number
of stars in this region we considered a less restrictive criterium 
on the metallicity and extended the sample to objects having $\rm -3 < [Fe/H] < -1$.
We carefully excluded the subgiants from the sample using the
recent work of Charbonnel and Primas (2005). This work takes advantage 
of the HIPPARCOS parallaxes in determining the stellar evolutionary status.
All the objects considered in figures \ref{fig4} and \ref{fig5}
are dwarfs or turn-off stars and comply with our prescriptions on metallicity.
However the four objects having upper limits in lithium detection
and $\rm T_{eff}$ above 5900 K have $\rm -1.66 \, dex <[Fe/H]<-0.88 \, dex$. Only one
of them is a confirmed dwarf (see table \ref{tab5}). Nevertheless because of their high
effective temperature these stars, if subdwarfs,  
have not experienced any significant additive 
extension of their outer convection zone since the turn-off. 
Thus their surface lithium 
fraction is not expected to have changed since the main sequence.
The computations have been made with both tachocline
and microscopic diffusion or only microscopic diffusion. In any case a 
$^7$Li plateau is predicted above 5500 K. This plateau lies 
between 0.2 and 0.4 dex below the initial abundances. In the microscopic
diffusion models the abundances moreover decrease by 0.2 dex between 5700 K and 
the hotter end of the plateau near 6400 K. 
On the plateau the main effect of the tachocline mixing is to correct this
effective temperature dependence : only a small decrease of less than 0.1 dex in 
$^7$Li remains towards the hot edge of the plateau.
These computations are in agreement with the predictions of Richard et al. (2005):
on the one hand the microscopic diffusion plays the major role in determining
the level of the plateau, on the other hand the shape of the plateau 
requires that a non standard mixing process occurs in the radiation zone.
For the sake of clarity the tachocline mixing models built
with the highest rotation velocity of the radiation zone (case D)
do not appear on figure \ref{fig4}. Above
6000 K these models do not show differences in $^7$Li exceeding
0.1 dex with the predictions of case A rotation models. In the framework
of the tachocline mixing the most rapid
rotation rates inferred from HB fast rotators are thus 
unable to explain the lithium poor
stars near the turn off. We shall return to this question in
the next chapter.
Below 5700 K the tachocline mixing models provide an explanation
of the observed $^7$Li pattern. On the contrary the pure microscopic 
diffusion models do not predict any $^7$Li depletion around 5500-5400 K.
The tachocline mixing therefore improves the situation with respect
to pure microscopic diffusion models. However there are still 
two remaining issues. First the observed plateau still lies 0.2 dex below
the predictions. Second the abundances below 5500 K appear largely 
scattered at a given effective temperature.

\begin{figure}[Ht]
\centering
\includegraphics[angle=90,width=8cm]{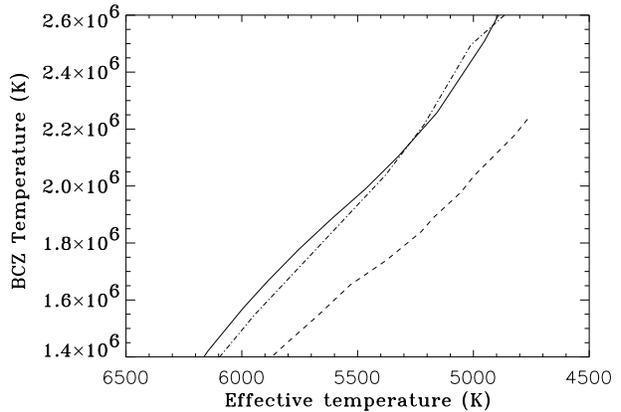}
\caption{$\rm T_{eff}-T_{BCZ}$ relation in the models of [Fe/H]=-2 dex, 
[Fe/H]=-3 dex and [Fe/H]=-2 
with the solar repartition among metals. Linestyle
conventions are the same as in figure \ref{fig5}.
Note the huge impact of metal repartition on the relation.}
\label{fig6} 
\end{figure}

\begin{figure*}[Ht]
\centering
\includegraphics[angle=90,width=8cm]{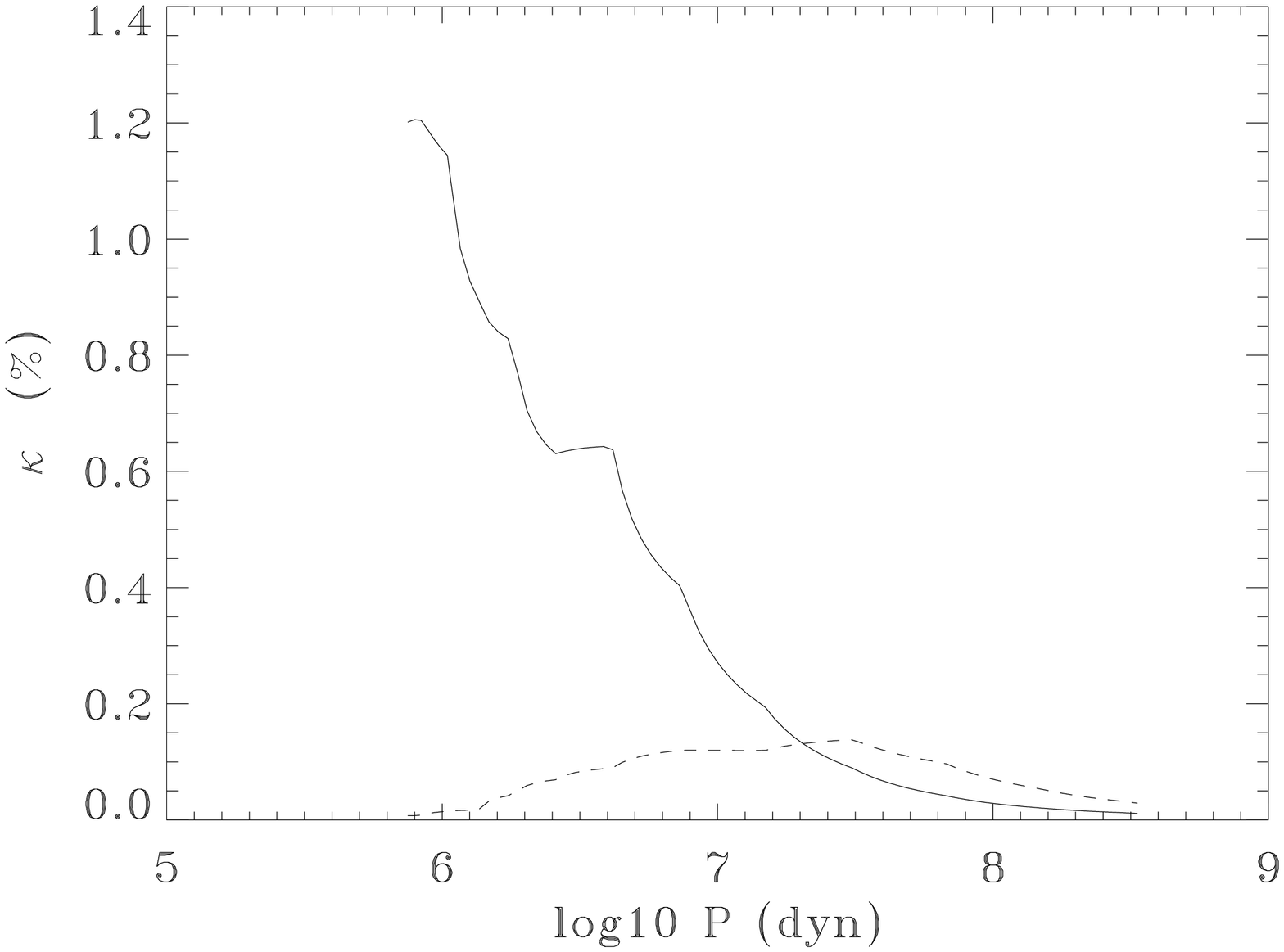}
\includegraphics[angle=90,width=8cm]{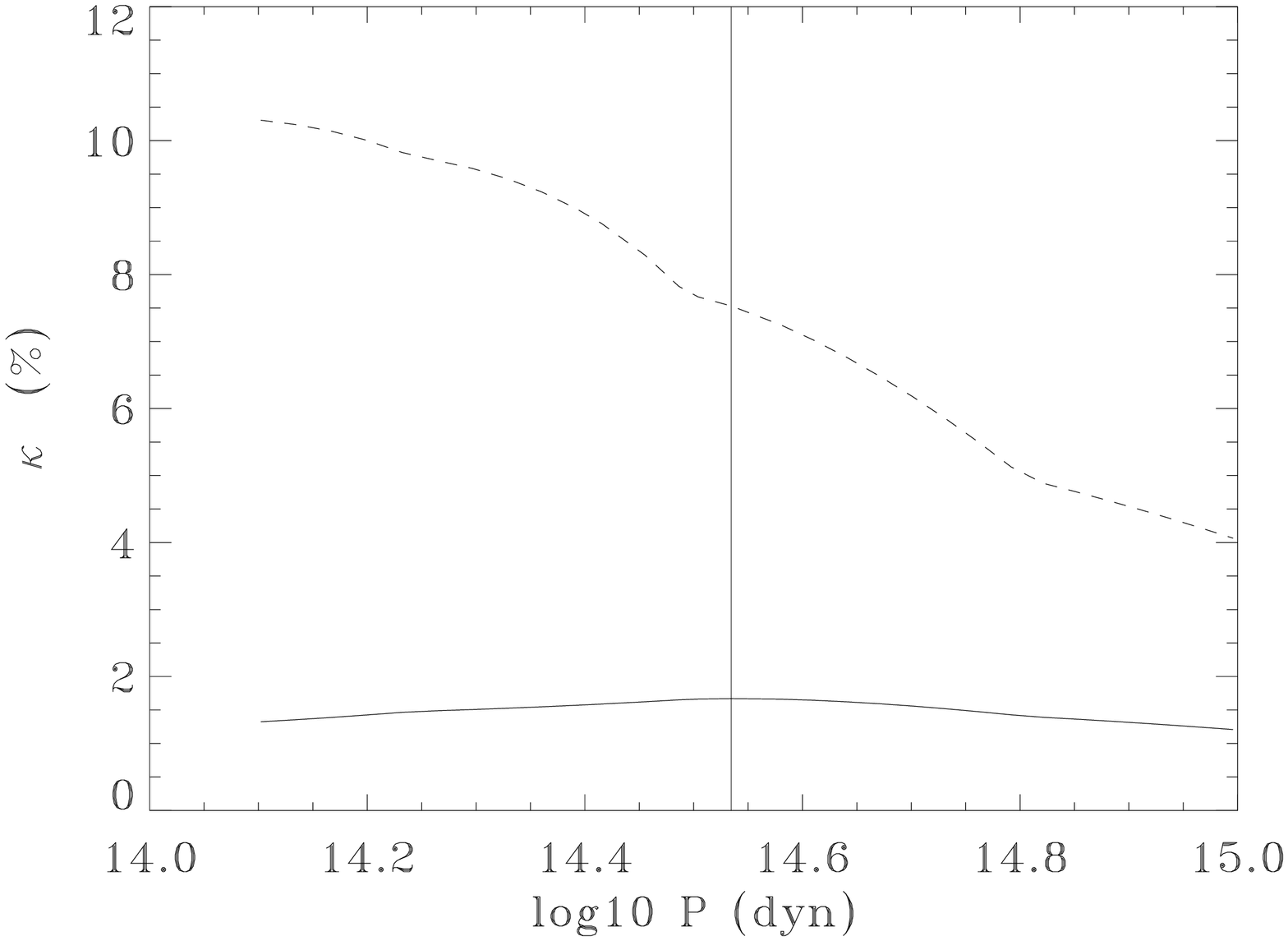}
\caption{Left panel: opacity contributions in \% for iron (solid line)
and CNO (dashed line) as function of the
pressure in the upper convection zone. The computations are for
the $\rm 0.62 M_{\odot}$ ($\rm T_{eff}=5260 K$) at 13 Gyr with [Fe/H]=-2 dex and
halo repartition among metals. The superadiabatic region extends down to
log P=7. Right panel : opacity contributions for iron 
and oxygen as a function of the
pressure in the lower convection zone. The linestyles are similar 
to the upper panel. The vertical line shows the BCZ.}
\label{fig7} 
\end{figure*}

A possible explanation of the discrepancy between the observed and
predicted plateau may stem from a slight underestimation of $^7$Li.
It has been suggested recently by Melendez \& Ramirez (2004) that 
such an underestimate could be related to an incorrect evaluation 
of the $\rm T_{eff}$ scale. The $\rm ^7Li$ abundance infered by these authors 
is 2.37 dex which is in perfect agreement with our predictions as shown on figure \ref{fig5}.
These observational results however still have to be confirmed.
What is the origin of the scatter in lithium abundances at a given $\rm T_{eff}$?
on the cooler side of the plateau?
As we have seen in \S \ref{sec4}, metallicity effects and
pre MS structural evolution are unable to explain this 
scatter. If one then investigates the effects of a MS rotational
mixing process it is tempting to relate the lithium scatter to different
rotation  histories. However, several clues coming from population I 
observations lead us to think that this is an incorrect interpretation. 
For instance, the Hyades solar-like stars all are slow rotators with $\rm v sin i < 10 km.s^{-1}$.
They presumably have experienced very different rotation histories
as is suggested by the observations of younger clusters. However
they show almost no scatter of their lithium abundances at a given temperature 
between 5000 and 6000 K (see for instance fig 6 of Thorburn et al. 1993).
Therefore in the case of the Hyades it appears artificial to relate differences of initial angular momentum
and differences of lithium abundances. One could imagine that the Hyades solar analogs
have kept radiative cores rotating with various speeds from their early evolution and
that the subsequent loss of angular momentum will induce different amounts of mixing
and surface lithium destruction. However the presence of fast rotating cores and slow
rotating envelopes after the Hyades age is not supported by helioseismic or 
early rotation history constraints. As we mentioned
in chapter \ref{sec3}, solid body rotation is probably achieved at the Hyades age.
If this is correct, the tachocline mixing (or any other rotation induced mixing)
combined with the microscopic diffusion will produce similar effects on lithium
during the subsequent evolution.
We suggest the lithium scatter originates in differences of the metal content instead.
The relevant quantity is not [Fe/H] however: as shown
on figure \ref{fig5} tachocline models with [Fe/H]=-2 dex or [Fe/H]=-3 dex exhibit
the same depletion pattern. The abundance repartition among metals 
has a much stronger influence on lithium. {\it The models keeping [Fe/H]=-2 dex
but with solar repartition among metals are less lithium 
depleted for any given effective temperature than the models assuming 
a population II metal distribution.} 
Changing the oxygen to iron
ratio by 0.3 dex affects the $^7$Li content more at a given effective
temperature than changing the global metallicity by 1 dex.
Figure \ref{fig6} displays 
the BCZ temperatures as a function of the effective temperatures for 
the three compositions displayed on figure \ref{fig5}. The relation is more affected
by a change in the metal ratios than by a change in the total 
metal fraction. 
The solar metal repartition models exhibit considerably lower BCZ temperatures
at any given effective temperature.
The reason for this behavior is twofold. The iron tunes the opacity 
near the top part of the convection zone whereas the
oxygen tunes the opacity near the base of the convection zone.
Figure \ref{fig7} illustrates these opposite behaviors
for a model where $^7$Li depletion becomes 
significant ($\rm T_{eff}=5270 K$).
An increase of the iron fraction tends to increase the
thermal gradient $\rm \nabla = \frac{dlnT}{dlnP}$ in the superadiabatic
region and thus tends to increase the entropy of the deep convection zone.
In turn this decreases the depth of the convection zone.
If the iron content is fixed so is the entropy of the convection
zone and the position of the BCZ then becomes
very sensitive to the opacity in this region. The oxygen
being the main opacity contributor near the BCZ, a diminution
of the oxygen content at fixed [Fe/H] will induce a drop
of BCZ temperature as illustrated on figure \ref{fig6}. The oxygen to iron fraction
drops by a factor 2 from the population II to the population I 
metals repartition, this in turn translates into a smaller 
$^7$Li depletion (figure \ref{fig5}).

{\it Because of the combined opacity roles of oxygen and iron 
we predict a correlation between $^7$Li abundance and [Fe/O].}
We expect this correlation only below 5500 K and moreover
for objects having [Fe/H] above $\sim$-2 dex. 
If the $\rm T_{eff}$ condition is not fulfilled
the BCZ is not close enough to the regions where
$^7$Li nuclear burning can occur. Alternatively if
the [Fe/H] condition is not fulfilled the metals
do not have enough impact on the stellar structure
to induce differences in the history of $^7$Li.
In this respect the anti-correlation between [Fe/O] and 
$^7$Li observed in NGC6752 by Pasquini et al. (2005)
does not contradict our expectations because it concerns
turn-off stars ($\rm T_{eff}>5900 K$). Following these authors we think that
the chemical properties of these objects are related to the material 
that formed them.
The case of a recently studied star \object{HE1327-2326} (Frebel et al. 2005) also drew 
our attention. With an effective temperature of $6180\pm 80$K,
\object{HE1327-2326} should lie on the lithium plateau. It is 
currently the lowest metallicity star known: [Fe/H]=-5.4 dex. However it is found to be
anomalously lithium poor, the upper limit being [$^7$Li]=1.6 dex.
The metal distribution in this object differs significantly from other population 
II dwarfs : e.g. $\rm [O/Fe]$ could be as high as 3.7 dex for this star and [C/Fe]=3.9 dex.
Considering the composition observed in this
star\footnote{The helium mass fraction was assumed 0.2479 while the oxygen 
and all the metals not mentioned in Frebel et al. (2005) were set to [X/Fe]=3.7 dex.}
we have tested the tachocline mixing model for $\rm T_{eff}=6200$ K.
Despite a large oxygen fraction, we found no significant 
depletion of $^7$Li at 13 Gyr. Because of the very small metallicity
the effects of the metals are quite moderate
on \object{HE1327-2326} even though [O/Fe] is huge.
This illustrate the [Fe/H] condition we mentionned just before.
In this respect it is interesting to note that the heavy elements
mass fraction and [O/H] ratio we compute for \object{HE1327-2326} are 
$\rm Z=3.1\, 10^{-4}$ and -1.8 respectively which is comparable 
to our standard composition case where $\rm Z=3.0 \,10^{-4}$ and
[O/H]=-1.7 dex (see table \ref{tab1}).

\subsection{Below the $\rm ^7Li$ plateau}

About 7\% of the halo turn-off stars are lithium poor with
abundances below 1.7 dex. Some of 
these objects show peculiar chemical abundance patterns while some
others cannot be distinguished from the other 'normal' lithium abundance
stars. However the majority of these stars
clearly rotates faster than their Li-normal counter parts
(Ryan \& Elliott 2005). At first sight this point should 
once again suggest a connection between lithium and rotation effects.
However we do not think that these stars are lithium poor because
they reached the ZAMS with more angular momentum. 
The distribution of rotational periods in ZAMS open-clusters
such as IC2602 and IC2391 do not show a small fraction of stars,
say between 5 and 10 \%, having periods clearly shorter than
the bulk of other comparable objects. It rather seems that rotation periods are uniformly
distributed on the 0.2-10 days range (Barnes et al. 1999).
If the ZAMS rotation distribution was responsible for
lithium poor halo dwarfs at turn-off, one would expect more 
stars below the plateau and a 
continuous distribution of them as function of the lithium fraction.
This is not the case: hot lithium poor stars are rare
and there is an observational gap of objects between 
$\rm [^7Li]\approx2.1$ dex and the upper limit of 1.7 dex. 
Furthermore the Li-poor objects generally belong to binary systems.
Indeed three out of four Li-poor stars
of the Ryan et al. (2002) sample are confirmed binaries, the
secondary component being presumably a compact object.
This leads the authors to assume the higher rotation 
rates were achieved through angular momentum accretion accompanying
a mass transfert from the former primary component.
Several processes could account for the low
lithium fraction : the variation of rotation rate and angular momentum
of the accreting object may have induced enough internal 
mixing. Alternatively the accreted
matter may have been initially lithium poor and induced a subsequent 
dilution.

Our purpose here is to briefly check these two possibilities in the case where
one considers the tachocline diffusion as the rotation induced mixing process. 
Table \ref{tab5} shows the Ryan et al. (2002) data for their Li-poor stars while
the table \ref{tab6} provides the general characteristics of our models computed
for [Fe/H]=-1 dex, more or less the metallicity of the observed objects. 
Out of the four objects of table \ref{tab5},
CD-3119466 might be peculiar: it is neither a confirmed 
fast rotator nor a spectroscopic binary
and is further away from turn-off than the other stars. On this basis we exclude it from
the subsequent discussion. Table \ref{tab6}
shows that the masses of the observed Li-poor stars probably
lie between 0.8 and $\rm 0.86 M_{\odot}$. Masses above 0.9 $\rm M_{\odot}$
are excluded
unless the stars are evolving towards the subgiant stage and already
have at least lost $\approx$ 160 K since the turn-off. 
Moreover such high masses would mean that the Li-poor objects have
half the age of the typical halo stars. Masses below
0.8 $\rm M_{\odot}$ also seem excluded because such objects should not reach the
high observed effective temperatures.

\begin{table*}[Ht]
  \begin{center}
    \caption{Turn off lithium poor objects. Data from Ryan et al. (2001) and Ryan et al. (2002).
These stars appear in figure \ref{fig2} and \ref{fig4}. Charbonnel \& Primas (2005) showed that Wolf 550 
is a dwarf and G202-65 is a subgiant. The two other stars are not mentioned by this work.}\vspace{1em}
    \renewcommand{\arraystretch}{1.2}
    \begin{tabular}[h]{lccccccc}
      Object        &  $\rm v sin i$   & $\rm T_{eff}$&   [Fe/H]       &  $^7$Li       \\
                    &  ($\rm km.s^{-1}$)    & (K)          &   (dex)        &  (dex)        \\
      \hline  									      
      CD-31 19466   &  $\rm <2.2$      & 5986         &   -1.66        & $\rm <1.49$   \\
      \hline									      
      Wolf 550      &  $\rm 5.5\pm0.6$ & 6269         &   -1.56        & $\rm <1.61$   \\
      \hline									      
      BD +51 1817   &  $\rm 7.6\pm0.3$ & 6345         &   -0.88        & $\rm <1.64$   \\
      \hline									      
      G202-65       &  $\rm 8.3\pm0.4$ & 6390         &   -1.32        & $\rm <1.67$   \\
      \hline
      \end{tabular}
   \label{tab5}
  \end{center}
\end{table*}

\begin{table*}[Ht]
  \begin{center}
    \caption{[Fe/H]=-1dex models characteristics at their maximum MS effective temperature.}\vspace{1em}
    \renewcommand{\arraystretch}{1.2}
    \begin{tabular}[h]{lcccccc}
      Mass           &  $\rm T_{eff}$   & age     &   Mass of the convection zone       \\
      ($\rm M_{\odot}$)  &  (K)             & (Gyr)   &   ($\rm M_{\odot}$)                     \\
      \hline  
         0.80        &  6233            & 12.15   &    $6.5\,10^{-3}$    \\
      \hline  
         0.83        &  6326            & 10.25   &    $3.6\,10^{-3}$    \\
      \hline
         0.86        &  6426            &  8.73   &    $1.9\,10^{-3}$    \\
      \hline
         0.90        &  6548            &  7.10   &    $7\,10^{-4}$    \\
      \hline
      \end{tabular}
   \label{tab6}
  \end{center}
\end{table*}

Now that the plausible mass range for lithium poor objects is established
let us consider the effect of the tachocline mixing. Ryan et al. (2002)
estimate a typical mass increase of $10^{-2} \rm M_{\odot}$. Thus the masses of 
the lithium poor progenitors were probably well above 0.75 $\rm M_{\odot}$ ($\rm T_{eff}=5700 K$ on
the ZAMS). Prior to the mass transfert these stars were hot
enough to be on the plateau and to have a 'normal' $^7$Li fraction.
The depletion impact of the tachocline mixing on a 0.8 $\rm M_{\odot}$ star
is negligible even for a high angular rotation velocity.
We built a 0.8 $\rm M_{\odot}$ model with constant rotation rate 
$\rm 1.27\,10^{-5} rad.s^{-1}$ on the MS.
This rate is chosen such that at 13 Gyr the surface equatorial 
velocity is 10 $\rm km.s^{-1}$ which is the order of magnitude of the 
velocity suggested by the vsini measurements (table \ref{tab5}).
However this model is extreme in terms of rotation and tachocline mixing 
because it keeps a high angular velocity all along its evolution. We find that
the final $^7$Li of this model is 2.32 dex. Despite the high rotation rate
no depletion occurs because the region 
of nuclear lithium burning lies too deep below the base of the convection
zone. 

Let us now consider the effect of lithium dilution.
As can be seen in table \ref{tab6} the masses of the outer convection 
zones of the considered objects are quite small. If the total mass of the
$^7$Li object is close to 0.8 $\rm M_{\odot}$ then a $3\,10^{-2} \rm M_{\odot}$ 
mass transfert of lithium free material would be necessary to dilute the $^7$Li below 
the observed upper limits and down to [$^7$Li]=1.61 dex. 
Alternatively if the mass is close to 0.86 $\rm M_{\odot}$ then a 
$10^{-2} \rm M_{\odot}$ accreted mass would be enough to decrease 
lithium to [$^7$Li]=1.55 dex.
Therefore the dilution mechanism of $^7$Li can explain the low
lithium surface abundances observed in some turn-off halo dwarfs. 
The need for an additive mixing process is required
only if the accreted mass is below $\rm 10^{-3} M_{\odot}$. Then
the decrease in surface lithium fraction would be less than 0.2 dex
even in the lighter case we estimate for the convection zone.
We note furthermore that in this case the process responsible for 
the mixing could not possibly be the tachocline mixing.
Abundance measurements of other elements than $^7$Li would certainly clarify
the issue of the origin of near turn-off $^7$Li deficient stars. 
For instance the detection of higher nitrogen and
lower carbon fractions than in the 'normal' halo dwarfs would support the
accretion scenario from a red giant star. The conversion of carbon into nitrogen is
ascribed to deep mixing and
is clearly observed from moderately (Smith, Briley
\& Harbeck 2005) to extremely (Spite et al. 2005) metal poor giants.
The presence of $^6$Li in these otherwise $^7$Li poor objects would also favour
the accretion scenario because it is difficult to imagine how
this more fragile isotope could survive if $^7$Li experienced
enhanced proton capture.

\subsection{Predictions on $\rm ^6Li$ }

\begin{figure}[Ht]
\centering
\includegraphics[angle=90,width=8cm]{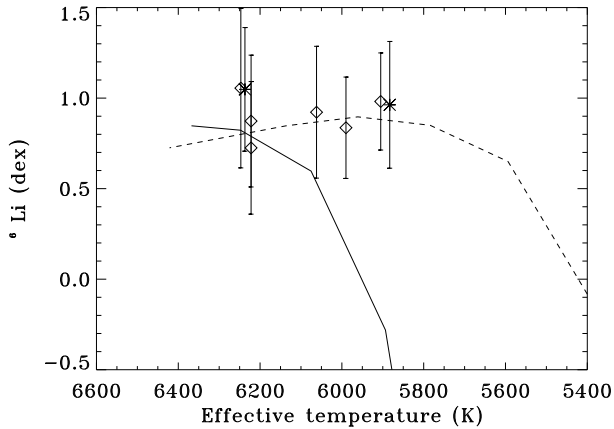}
\caption{$\rm T_{eff}-^6Li$ relation for tachocline diffusion models of
buoyancy frequency 10$\mu$ Hz. Solid line: [Fe/H]=-2 dex and halo repartition 
among metals. Dashed line [Fe/H]=-2 dex and solar repartition 
among metals. The data and error bars are from Asplund et al. (2005a) (diamonds)
and Nissen et al. (1999) (star symbols).}
\label{fig8} 
\end{figure}

The region of nuclear destruction of $^6$Li is close to the lower limit 
of the outer convection zone in halo dwarf stars. 
In this respect $^6$Li provides tighter constraints than $^7$Li on these
objects. Figure \ref{fig8} is the equivalent for $^6$Li to figure \ref{fig4} for
$^7$Li. The data are those mentionned in \S \ref{sec4}. We stress that 
not all the stars are confirmed as dwarfs by the recent analysis of
Charbonnel \& Primas (2005). However the subgiants of the sample 
are all hot enough that
no significant dilution or nuclear destruction of $^6$Li is expected
from deeper convective zones. Two tachocline diffusion models are displayed for [Fe/H]=-2
and halo star metal repartition and for [Fe/H]=-2 and solar metal repartition.
Above 6000 K the agreement with the data is good for the two sets of models.
Because the turbulence induced by tachocline mixing is shallow,
both models brake diffusion of heavy elements around the turn-off
without inducing significant $^6$Li destruction here. This
is an improvement with respect to the computations of Richard et al. (2005).
Nonetheless below 6000 K the tachocline models for the composition expected 
for halo stars predicts too strong a depletion. 
The presence of $^6$Li in stars as cool as 5900 K
appears problematic. However no hasty conclusion should
be drawn because of the complexity of the situation.
First, non detection stars are not represented on the plot.
We recall that for the Asplund et al. (2005a) sample
only 9 out of 24 stars yielded to an unambiguous $^6$Li detection.
Second because of the still rare $^6$Li detections, the data of 
stars having $\rm -2 < [Fe/H] < -1$ dex are compared to the
models. For the sake of simplicity we 
consider a unique $^6$Li initial fraction ($\approx$ 1.1 dex after 
correction for diffusion effects). However
cosmic ray activity predicts a significant increase
of $^6$Li over the actual metallicity range (Rollinde et al. 2005).
Finally we expect that the
metal repartition effects will change the effective temperature/BCZ
temperature relation as aforementioned.
In the same manner as in the preceding chapter
we considered the tachocline mixing impact when
[Fe/H]=-2 dex and the metal repartition is solar. As can be seen
on figure \ref{fig8} the agreement to the data is then quite
improved. This suggest there should be a [Fe/O] to $^6$Li
correlation for halo dwarfs around the turn-off.

\section{Summary}\label{sec6}

We examined in detail the evolution of both lithium isotopes in halo stars of effective
temperatures in the 4700 to 6500 K range. Their depletion history has been
followed during the pre main sequence and the main sequence up to the age of 13 Gyr
and compared with observations. The $^9$Be history was also followed
but no significant evolution of this species is predicted because of
its higher temperature for proton capture compared to lithium.
Our models include the most recent physics (\S \ref{sec2}) in terms of 
equation of state and opacities (OPAL tables),
nuclear reaction rates (NACRE compilation) and atmospheric boundary 
conditions (NextGen atmosphere models). Special attention
was given to the metal fractions impact through opacity effects: models
with [Fe/H]=-3, -2, -1 and -0.7 dex have been considered. In all cases the
helium mass fraction and $^7$Li number fraction were
assumed to be primordial: Y=0.2479 and [$^7$Li]=2.62 dex 
respectively (Coc et al. 2004). The initial $^6$Li number fraction
was assumed to be 1.13 dex. The microscopic 
diffusion has always been taken into account. Moreover we have studied 
the effects of the tachocline turbulent diffusion, a rotationally induced
mixing that was previously demonstrated to explain lithium and beryllium
history in the Sun and other population I stars. The two free parameters 
of the tachocline diffusion process are calibrated on the Sun (\S \ref{sec3}). 
The rotation history of a 0.7 $\rm M_{\odot}$ model is used to account for 
the tachocline mixing. This rotation history was computed in detail,
taking into account the structural changes and surface angular momentum 
losses (\S \ref{sec3}).

During the pre main sequence all the models evolve from a fully 
convective stage to a radiative core / convective envelope structure.
Meanwhile the temperature at the boundary between these two regions 
reaches a maximum. The peak temperature increases as
the mass of the star decreases which results in an enhanced lithium depletion
on the ZAMS for cooler objects. As is well known, the observed
$^7$Li abundance pattern shows a plateau
in $\rm ^7Li \, vs \, T_{eff}$
from the turn-off down to $\rm T_{eff} \approx$ 5500 K
while below this value the depletion gradually increases with decreasing 
effective temperature. We showed that this pattern cannot 
be explained by pre main sequence depletion regardless of 
metallicity below [Fe/H]=-1 dex (\S \ref{sec4}). The predictions of depletion are
at least 0.7 dex above the observations in the effective temperature 
range between 5500 K and 5000 K. Thus an additive non standard mixing process
in the radiation zone is necessary to explain the present day observations.
It is noteworthy that, contrary to the population I case, the ZAMS predicted 
$\rm T_{eff}-^7Li$ relation does not depend on the metal content. $^6$Li abundance
provide additional informations about the pre main sequence evolution. 
The expected depletion of this isotope depends on the metallicity and clearly exceeds 
what is observed in some halo dwarfs. This issue is similar to what we found
in earlier work for population I star with respect to $^7$Li pre main sequence
depletion (PTC02). For surface temperatures cooler than 6000 K
some stars are observed on their late main sequence with a $^6$Li
content at least 1.5 dex above pre main sequence predictions. 
This indicates that the initial conditions or the 
early structural changes of population II
stars are not understood.

The pre main sequence results show that a non-standard main sequence
mixing near the base of the convection zone
 is necessary to explain the $^7$Li fraction below $\rm T_{eff}=5500$ K.
The $\rm T_{eff}-^7Li$ relation calculated with microscopic diffusion 
models at 13 Gyr also suggest that a turbulent process prevents too efficient microscopic 
diffusion near the turn-off. The tachocline mixing provides 
good results for the $^7$Li abundance pattern on the
cool side of the plateau. Moreover as the phenomenon creates turbulence below the
convection zone it also prevents increased depletion where the microscopic
diffusion timescale becomes shorter. Consequently the Spite plateau is well
reproduced (\S \ref{sec5}). Its average level lies between 2.4 and 2.3 dex. This is slightly
above the bulk of the observations, but in perfect agreement with the 
recent results of Melendez \& Ramirez (2004). We showed that the early main sequence rotation 
history has no impact on $^7$Li provided that nearly solid body rotation is efficiently enforced
in low mass stars. There are indeed several observational clues that the
solid body rotation is rapidly achieved in low mass stars. Furthermore using constraints from horizontal branch 
stars rotation we showed that even if halo dwarfs keep 
fast rotating cores this should not induce increased lithium depletion
around turn-off. The tachocline mixing requires a few gigayears to become effective
whereas solid body rotating stars achieve similar rotation rates 
after $\approx$ 0.2 Gyr. Consequently the small scatter on the $^7$Li plateau 
is not contradictory with scattered rotational rates on the ZAMS
even if one considers that the turbulence below the convection zone 
is connected to rotation. In the framework of the tachocline mixing the
initial angular momentum and the surface abundances are not correlated.
The tachocline mixing is robust with respect to the new
observations. It predicts $^6$Li abundances in agreement 
with the observations above 6000 K.
The rare objects exhibiting a $^7$Li abundance
clearly below the plateau and an effective temperature above 6000 K
cannot be explained in the framework of the tachocline process
despite their abnormaly high rotation rates. We show however that
because of their light convection zones, accretion of lithium free matter below
a few percent of solar mass could explain the actual observations.
The detection of $^6$Li or enhanced nitrogen content
in these objects would be strong support in favor of such an 
accretion scenario from an evolved star.
We suggest that the scatter in $^7$Li for stars cooler than the red edge of
the Spite plateau is related to differences in the metal repartition 
in the stars instead
of different rotation histories. The variation of the metals ratio 
from population II repartition to population I repartition {\it at fixed $T_{eff}$}
significantly diminishes lithium depletion. We predict a correlation 
between [Fe/O] and lithium abundance in both
isotopes at any given effective temperature below 5500 K for $^7$Li
and 6000 K for $^6$Li. These correlations should be observed 
in dwarfs stars where the metal repartition goes from halo 
repartition to disk repartition (i.e. around [Fe/H]=-1.5 dex). 
To our knowledge
such correlations have not yet been investigated.


\begin{thebibliography}{}

\bibitem{1}Akerman, C.J.,Carigi, L., Nissen, P.E., Pettini, M., \& Asplund, M., 2004 A\&A, 414, 931
\bibitem{1}Alexander, D. R., Ferguson, J. W., 1994, ApJ, 437, 879
\bibitem{1}Angulo, C., et al., Nucl. Phys. A656 (1999)3-187
\bibitem{1}Asplund, M., Nissen, P. E., Lambert, D. L., Primas, F., Smith, V. V., 2005a, Proceedings of IAU symposium 228, Ed. V. Hill, P. Fran\c{c}ois, F. Primas.
\bibitem{1}Asplund, M., Nissen, P. E., Lambert, D. L., Primas, F., Smith, V. V., 2005b, submitted to ApJ
\bibitem{1}Bahcall, J. N., Pinsonneault, M. H., Wasserburg, G. J., 1995, RvMP, 67, 781
\bibitem{1}Barnes, S. A., Sofia, S., Prosser, C. F., Stauffer, J. R., 1999, ApJ, 516, 263
\bibitem{1}Behr, B. B., 2003, ApJSS, 149,67
\bibitem{1}B\"{o}hm-Vitense, E., 1958, Zs. f. Ap., 46, 108
\bibitem{1}Bonifacio, P., Molaro, P., 1997, MNRAS, 285, 847
\bibitem{1}Boesgaard A. M., Deliyannis, C. P., Stephens, A., King, J. R., 1998, ApJ 493, 206
\bibitem{1}Bouvier, J., Forestini, M., Allain, S., (BFA97) 1997, A\&A, 326, 1023
\bibitem{1}Brun, A. S., Turck-Chi\`eze, S., Zahn, J. P. 1999, ApJ, 525, 1032
\bibitem{1}Brun, A. S.; Antia, H. M., Chitre, S. M., Zahn, J.-P., 2002, ApJ, 391, 725
\bibitem{1}Burles, S., 2002, P\&SS, 50, 1245
\bibitem{1}Cayrel, R., Spite, M., Spite, F., Vangioni-Flam, E., Cassé, M., Audouze, J., 1999, A\&A, 343, 923
\bibitem{1}Cayrel, R., Depagne, E., Spite, M., Hill, V., Spite, F., Fran\c{c}ois, P., Plez, B., Beers, T., Primas, F., Andersen, J., and 4 coauthors, 2004, A\&A, 416, 1117
\bibitem{1}Charbonnel, C., Primas, F., 2005, A\&A, accepted
\bibitem{1}Cyburt, R. H., Fields, B. D., Olive, K. A., 2004, Phys. Rev. D, 69, 123519 
\bibitem{1}Cyburt, R. H., Fields, B. D., Olive, K. A., 2003, Phys. Lett. B, 567, 227
\bibitem{1}Donahue, R. A., Saar, S. H., Baliunas, S. L., 1996, ApJ, 466, 384
\bibitem{1}Frebel, A., Aoki, W., Christlieb, N., Ando, H., Asplund, M., Barklem, P. S., Beers, T. C., Eriksson, K., Fechner, C., Fujimoto, M. Y., and 9 coauthors, 2005, Nature,434, 871
\bibitem{1}Gratton, R. G., 1989, A\&A, 208, 171
\bibitem{1}Hobbs, L. M., Duncan, D. K., 1987, ApJ,  317, 796
\bibitem{1}Iglesias, C. A., \& Rogers, F., J., 1996, ApJ, 464, 943
\bibitem{1}Magain, P., 1989, A\&A, 209, 211
\bibitem{1}Melendez, J., Ramirez, Y, 2004, ApJ, 615, 33
\bibitem{1}Michaud, G., Proffitt, C. R., 1993, Inside the stars , IAU colloquium, 137, ASP Conference series, eds. A. Baglin, \& W. W. Weiss vol 40, 426
\bibitem{1}Montalb\'an, J. \& Schatzman, E., 2000, A\&A 354, 943
\bibitem{1}Morel, P., 1997, A\&AS, 124, 597
\bibitem{1}Nissen, P. E., Lambert, D. L., Primas, F., Smith, V. V., 1999, A\&A, 348, 211
\bibitem{1}Pasquini, L., Bonifacio, P., Molaro, P., Francois, P., Spite, F., Gratton, R. G., Carretta, E., Wolff, B., 2005, A\&A, 441, 549
\bibitem{1}Piau, L., \& Turck-Chi\`eze, S., (PTC02) 2002, ApJ, 566, 419
\bibitem{1}Piau, L., Randich, S., \& Palla, F., 2003, A\&A, 408, 1037
\bibitem{1}Piau, L., Ballot, J., \& , Turck-Chi\`eze, S., 2005, A\&A, 430, 571
\bibitem{1}Proffitt, C. R., Michaud, G., 1991, ApJ, 371, 584
\bibitem{1}Reeves, H., Fowler, W. A., Hoyle, F., 1970, Nature, 226, 727
\bibitem{1}Rogers, F. J., Nayfonov, A., 2002, ApJ, 576, 1074
\bibitem{1}Ryan, S. G., Beers, T. C., Deliyannis, C. P., Thorburn, J. A., 1996, ApJ 458, 543
\bibitem{1}Ryan, S. G., Deliyannis, C. P., 1998, ApJ, 500, 398
\bibitem{1}Ryan, S. G., Norris, J. E., Beers, T. C., 1999, ApJ, 523, 654
\bibitem{1}Ryan, S. G.; Kajino, T., Beers, T. C., Suzuki, T. K., Romano, D., Matteucci, F. Rosolankova, K., 2001, ApJ 549, 55
\bibitem{1}Ryan et al. 2002, ApJ, 571, 501
\bibitem{1}Smith, V. V., Lambert, D. L., Nissen, P. E., 1998, ApJ, 506, 405
\bibitem{1}Smith, G. H., Briley, M. M., Harbeck, D., 2005, AJ, 129, 1589
\bibitem{1}Spiegel, E. A., Zahn, J.P., 1992, A\&A, 265, 106
\bibitem{1}Spite, F., Spite, M., 1982, A\&A 115, 457
\bibitem{1}Spite, F., Spite, M., 1993, A\&A 279, L9
\bibitem{1}Spite, M., Francois, P., Nissen, P. E., Spite, F., 1996, A\&A 307, 172
\bibitem{1}Spite, M., Cayrel, R., Plez, B., Hill, V., Spite, F., Depagne, E., François, P., Bonifacio, P., Barbuy, B., Beers, T., and 4 coauthors, 2005, A\&A, 430, 655
\bibitem{1}Talon, S., Kumar, P., Zahn, J. P., 2002, ApJ, 574, 175
\bibitem{1}Talon, S., Charbonnel, C., 2004, A\&A, 418, 1051
\bibitem{1}Thorburn, J. A., Hobbs, L. M., Deliyannis, C. P., Pinsonneault, M. H., 1993, ApJ, 415, 150
\bibitem{1}Thorburn, J. A., 1994, ApJ, 421, 318
\bibitem{1}Thompson, M. J., Christensen-Dalsgaard, J., Miesch, M. S., Toomre, J., 2003, A\&ARA, 41, 599
\bibitem{1}VandenBerg, D., 2000, ApJS, 129, 315
\bibitem{1}Vangioni-Flam, E., Cass\'e, M., Cayrel, R., Audouze, J., Spite, M., Spite, F., 1999, NewA, 4, 245
\bibitem{1}Ventura, P. Zeppieri, A., Mazzitelli, I., D'Antona, F., 1998, A\&A, 331, 1011
\bibitem{1}Zahn, J. P. , 2004, 'Hydrodynamic models of the tachocline', Cambridge University press. 

\end{thebibliography}
\end{document}